\documentclass[conference]{IEEEtran}
\IEEEoverridecommandlockouts

\usepackage{tikz}
\usepackage{amssymb}
\usepackage{amsthm}
\usepackage{amsmath}
\usepackage{mathtools}
\usepackage{framed}
\usepackage{balance}
\usepackage{subcaption}
\usepackage[caption=false,font=scriptsize, labelfont=sf,textfont=sf]{subfig} 
\usepackage{pifont}
\usepackage{enumitem}
\usepackage{booktabs}
\usepackage{multirow}
\usepackage{pgfplots}
\usepackage{graphicx}
\usepackage{glossaries}
\usepackage{array}
\usepackage{comment}
\usepackage{multirow}
\usepackage{makecell}
\usepackage{cite}
\usetikzlibrary{shapes}
\usepackage{float}
\usepackage{stfloats}
\usepackage[capitalise]{cleveref}
\usepackage{verbatim}
\usepackage{textcomp}
\usepackage{titlesec}
\usepackage[table]{xcolor}
\usepackage{tabstackengine}
\usepackage{booktabs,array}
\usepackage{stackengine}
\usepackage{adjustbox}
\usepackage{mathrsfs}

\renewcommand{\theequation}{{\color{blue}\arabic{equation}}}

\newcommand{\red}[1]{\textcolor{red}{#1}}

\newtheorem{definition}{Procedure}
\let\olddefinition\definition
\renewcommand{\definition}{\olddefinition\normalfont}

\newtheoremstyle{myremark}
  {\topsep}
  {\topsep}
  {\itshape}
  {0pt}
  {\scshape}
  {.}
  { }
  {}

\theoremstyle{myremark}
\newtheorem{remark}{Remark}

\makeatletter
\let\save@mathaccent\mathaccent
\newcommand*\if@single[3]{%
  \setbox0\hbox{${\mathaccent"0362{#1}}^H$}%
  \setbox2\hbox{${\mathaccent"0362{\kern0pt#1}}^H$}%
  \ifdim\ht0=\ht2 #3\else #2\fi
  }
\newcommand*\rel@kern[1]{\kern#1\dimexpr\macc@kerna}
\newcommand*\wideaccent[2]{\@ifnextchar^{{\wide@accent{#1}{#2}{0}}}{\wide@accent{#1}{#2}{1}}}
\newcommand*\wide@accent[3]{\if@single{#2}{\wide@accent@{#1}{#2}{#3}{1}}{\wide@accent@{#1}{#2}{#3}{2}}}
\newcommand*\wide@accent@[4]{%
  \begingroup
  \def\mathaccent##1##2{%
    \let\mathaccent\save@mathaccent
    \if#42 \let\macc@nucleus\first@char \fi
    \setbox\z@\hbox{$\macc@style{\macc@nucleus}_{}$}%
    \setbox\tw@\hbox{$\macc@style{\macc@nucleus}{}_{}$}%
    \dimen@\wd\tw@
    \advance\dimen@-\wd\z@
    \divide\dimen@ 3
    \@tempdima\wd\tw@
    \advance\@tempdima-\scriptspace
    \divide\@tempdima 10
    \advance\dimen@-\@tempdima
    \ifdim\dimen@>\z@ \dimen@0pt\fi
    \rel@kern{0.6}\kern-\dimen@
    \if#41
      #1{\rel@kern{-0.6}\kern\dimen@\macc@nucleus\rel@kern{0.4}\kern\dimen@}%
      \advance\dimen@0.4\dimexpr\macc@kerna
      \let\final@kern#3%
      \ifdim\dimen@<\z@ \let\final@kern1\fi
      \if\final@kern1 \kern-\dimen@\fi
    \else
      #1{\rel@kern{-0.6}\kern\dimen@#2}%
    \fi
  }%
  \macc@depth\@ne
  \let\math@bgroup\@empty \let\math@egroup\macc@set@skewchar
  \mathsurround\z@ \frozen@everymath{\mathgroup\macc@group\relax}%
  \macc@set@skewchar\relax
  \let\mathaccentV\macc@nested@a
  \if#41
    \macc@nested@a\relax111{#2}%
  \else
    \def\gobble@till@marker##1\endmarker{}%
    \futurelet\first@char\gobble@till@marker#2\endmarker
    \ifcat\noexpand\first@char A\else
      \def\first@char{}%
    \fi
    \macc@nested@a\relax111{\first@char}%
  \fi
  \endgroup
}
\makeatother

\newcommand\widebar{\wideaccent\overline}

\makeatletter
\newcommand{\doublewidetilde}[1]{{%
  \mathpalette\double@widetilde{#1}%
}}
\newcommand{\double@widetilde}[2]{%
  \sbox\z@{$\m@th#1\widetilde{#2}$}%
  \ht\z@=.9\ht\z@
  \widetilde{\box\z@}%
}
\makeatother

\newcommand\scaleddot{\scalebox{.89}{.}}

\makeatletter
\renewcommand{\dddot}[1]{%
  {\mathop{\kern\z@#1}\limits^{\makebox[0pt][c]{\vbox to-2.2\ex@{\kern-\tw@\ex@
   \hbox{\normalfont\scaleddot\kern-0.5pt\scaleddot\kern-0.5pt\scaleddot}\vss}}}}}
\renewcommand{\ddddot}[1]{%
  {\mathop{\kern\z@#1}\limits^{\makebox[0pt][c]{\vbox to-2.2\ex@{\kern-\tw@\ex@
   \hbox{\normalfont\scaleddot\kern-0.5pt\scaleddot\kern-0.5pt\scaleddot\kern-0.5pt\scaleddot}\vss}}}}}
\makeatother

\makeglossaries

\newacronym{ISAC}{ISAC}{integrated sensing and communications}
\newacronym{BS}{BS}{base station}
\newacronym{RF}{RF}{radio-frequency}
\newacronym{DAC}{DAC}{digital-to-analog converter}
\newacronym{IRS}{IRS}{intelligent reflecting surface}
\newacronym{PAPR}{PAPR}{peak-to-average power ratio}
\newacronym{CSI}{CSI}{channel state information}

\newacronym{COTS}{COTS}{commercial-off-the-shelf}
\newacronym{ULA}{ULA}{uniform linear array}

\newacronym{RRM}{RRM}{radio resource management}
 
\newacronym{AWGN}{AWGN}{additive white Gaussian noise}
\newacronym{SNR}{SNR}{signal-to-noise ratio}
\newacronym{DPG}{DPG}{directional power gain}
\newacronym{DR}{DR}{data rate}
\newacronym{SINR}{SINR}{signal-to-interference-plus-noise ratio}
\newacronym{SDR}{SDR}{semidefinite relaxation}
\newacronym{SDP}{SDP}{semidefinite programming}
\newacronym{SCA}{SCA}{successive convex approximation}
 
\newacronym{MILP}{MILP}{mixed-integer linear program} 
\newacronym{MINLP}{MINLP}{mixed-integer nonlinear program} 
\newacronym{MISDP}{MISDP}{mixed-integer semidefinite program} 

\newacronym{AOA}{AOA}{angle of arrival}
\newacronym{AOD}{AOD}{angle of departure}
\newacronym{RC}{RC}{reflection coefficient}
\newacronym{SI}{SI}{self-interference}

\newacronym{LoS}{LoS}{line-of-sight}
\newacronym{NLoS}{NLoS}{non-LoS}

\newacronym{LHS}{LHS}{left-hand-side}
\newacronym{RHS}{RHS}{right-hand-side}

\newacronym{ES}{ES}{exhaustive search}
\newacronym{BnC}{BnC}{branch-and-cut}

\begin{document}



\title{\Huge Resilient Full-Duplex ISAC in the Face of Imperfect SI Cancellation: Globally \\ Optimal Timeslot Allocation and Beam Selection }


\author{
\IEEEauthorblockN{
Luis F. Abanto-Leon\IEEEauthorrefmark{2} 
and
Setareh Maghsudi\IEEEauthorrefmark{4},
}
\IEEEauthorblockA{
Ruhr University Bochum, Bochum, Germany\\
\IEEEauthorrefmark{2}Email: l.f.abanto@ieee.org
\quad
\IEEEauthorrefmark{4}Email: setareh.maghsudi@ruhr-uni-bochum.de}
}

\maketitle


\begin{abstract}
	This work addresses the \gls{RRM} design in downlink full-duplex \gls{ISAC} systems, jointly optimizing timeslot allocation and beam selection under imperfect self-interference cancellation. Timeslot allocation governs the distribution of discrete channel uses between sensing and communication tasks, while beam selection determines transmit and receive directions along with adaptive beamwidths. The joint design leads to a semi-infinite, nonconvex \gls{MINLP}, which is difficult to solve. To overcome this, we develop a tailored reformulation strategy that transforms the problem into a tractable \gls{MILP}, enabling globally optimal solutions. Our approach provides insights into the coordinated optimization of timeslot allocation and beam selection, enhancing the efficiency of full-duplex \gls{ISAC} systems while ensuring resilience against residual self-interference.
\end{abstract}




\glsresetall
\section{Introduction} \label{sec:introduction}

\Gls{ISAC} marks a groundbreaking advancement in wireless technology by seamlessly combining sensing and communications functionalities within the same hardware, spectrum, and waveform  \cite{ balef2023:adaptive-energy-efficient-waveform-design-joint-communication-sensing-multiobjective-multiarmed-bandits}. This tight integration promises to maximize radio resource utilization efficiency and reduce costs, enabling enhanced performance and capabilities across a wide range of applications.

Millimeter-wave and terahertz frequency bands are highly attractive for \gls{ISAC} due to their short wavelengths, which enable high resolution for precise sensing \cite{yao2014:terahertz-active-imaging-radar-preprocessing-experiment-results}, and their abundant spectrum availability, which supports high-throughput communications \cite{askar2024:mobilizing-terahertz-beam-d-band-analog-beamforming-front-end-prototyping-long-range-6g-trials}. Prior works, e.g., \cite{zhuo2024:multi-beam-integrated-sensing-communication-state-of-the-art-challenges-opportunities, abanto2024:hierarchical-functionality-prioritization-multicast-isac-optimal-admission-control-discrete-phase-beamforming}, have demonstrated the strong potential of combining \gls{ISAC} with high-frequency bands to address the stringent connectivity and sensing requirements of future wireless systems.

At high frequencies, effective beamforming design is essential to unlock the full potential of \gls{ISAC}. While digital beamforming offers high performance, its hardware complexity and costs scale unfavorably with frequency. This has driven growing interest in analog beamforming for \gls{ISAC}, particularly in practical beam selection strategies, due to its cost-effectiveness and simple implementation \cite{zhao2020:m-cube:-millimeter-wave-massive-mimo-software-radio}. However, most existing studies, with few exceptions (e.g., \cite{du2023:integrated-sensing-communications-v2i-networks-dynamic-predictive-beamforming-extended-vehicle-targets, zhang2023:robust-beamforming-design-uav-communications-integrated-sensing-communication}), considered beams with fixed beamwidth configurations. This assumption can be restrictive, as \gls{ISAC} systems must accommodate diverse user and target requirements, necessitating adaptive beamwidth control. Furthermore, prior \gls{ISAC} work predominantly focused on transmit-side beamforming, often overlooking the role of receive beamforming. \emph{This motivates the need for an  \gls{ISAC} \gls{RRM} design that jointly optimizes beam direction and beamwidth at both the transmitter and receiver.}

Time allocation is another critical aspect in analog beamforming systems. With only a single \gls{RF} chain, analog beamformers can process one signal at a time, requiring efficient time sharing to satisfy both sensing and communication demands. While time allocation has been extensively studied in the sensing literature, e.g., \cite{ zhang2023:fast-solver-dwell-time-allocation-phased-array-radar}, and in the communication literature, e.g., \cite{abanto2020:hydrawave-multi-group-multicast-hybrid-precoding-low-latency-scheduling-ubiquitous-industry-4-0-mmwave-communications , nguyen2017:joint-fractional-time-allocation-beamforming-downlink-multiuser-miso-systems }, only a limited number of studies have addressed this crucial aspect in the \gls{ISAC} context, e.g., \cite{xu2023:sensing-enhanced-secure-communication-joint-time-allocation-beamforming-design, wang2024:resource-allocation-isac-networks-application-target-tracking}. Notably, \cite{xu2023:sensing-enhanced-secure-communication-joint-time-allocation-beamforming-design} considered continuous-time allocation, whereas \cite{wang2024:resource-allocation-isac-networks-application-target-tracking} focused on its discrete-time counterpart, the latter being particularly relevant, as it aligns with the timeslot-based structure of modern communication systems. \emph{Therefore, a \gls{RRM} design that incorporates timeslots as a key resource is crucial for meeting heterogeneous sensing and communication requirements, especially under the physical constraints of analog beamforming.}


\begin{figure}[!t]
	\centering
	\includegraphics[width=0.98\columnwidth]{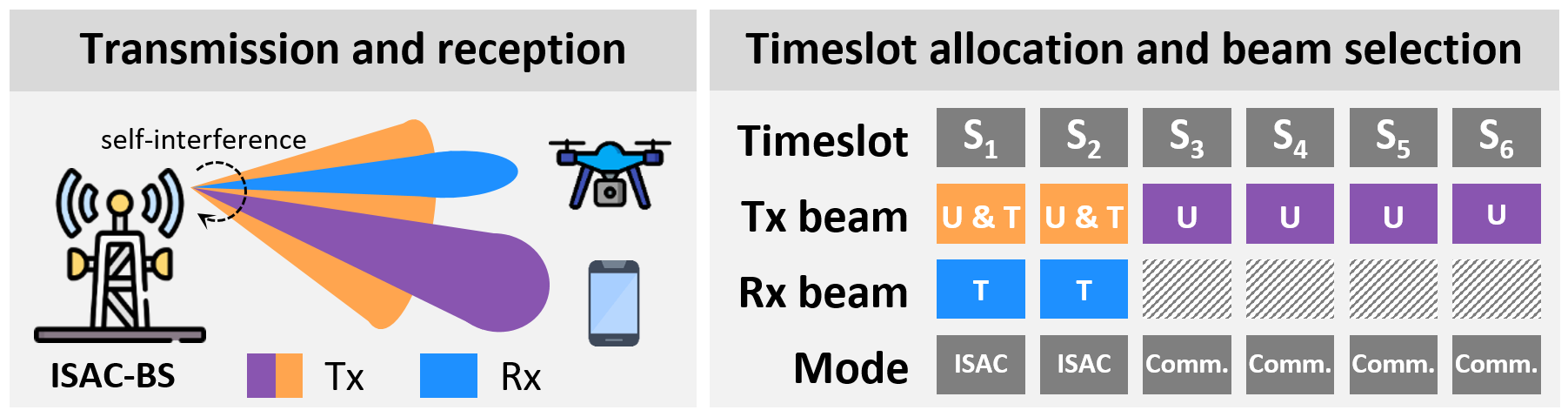}
	\caption{System model comprising a BS, a user, and a target.}	
	\label{fig:system-model}
	\vspace{-4mm}
\end{figure}

To further improve resource utilization, full-duplex sensing has emerged as a promising paradigm \cite{smida2023:full-duplex-wireless-6g-progress-brings-new-opportunities-challenges}. However, its performance is fundamentally limited by \gls{SI} \cite{hernangomez2024:optimized-detection-analog-beamforming-monostatic-integrated-sensing-communication}. While many existing studies assume either perfect \gls{SI} cancellation or full knowledge of the \gls{SI} channel, e.g., \cite{xu2025:robust-isac-transceiver-beamforming-design-under-low-resolution-ad-da-converters, jiang2025:full-duplex-isac-enabled-d2d-underlaid-cellular-networks-joint-transceiver-beamforming-power-allocation}, practical systems inevitably suffer from residual interference due to incomplete channel knowledge and hardware impairments. Such residual \gls{SI} can critically degrade both communication and sensing performance. \emph{Therefore, a \gls{RRM} design that explicitly accounts for imperfect \gls{SI} knowledge is essential to ensure resilient operation.}

Building on the above motivation, this paper investigates a novel \gls{RRM} problem for full-duplex \gls{ISAC} systems that jointly optimizes timeslot allocation and beam selection, encompassing transmit/receive directions and beamwidths, while explicitly incorporating imperfect knowledge of residual \gls{SI}. The resulting formulation yields a semi-infinite nonconvex \gls{MINLP}, which we systematically transform into an \gls{MILP} through a series of reformulations, enabling globally optimal solutions. Simulation results reveal insightful interdependencies in timeslot and beam sharing across functionalities, demonstrating how coordinated allocation can maximize overall resource efficiency.

\emph{Notation}: Boldface capital letters $ \mathbf{X} $ and lowercase letters $ \mathbf{x} $ denote matrices and vectors, respectively. The transpose and Hermitian transpose of $ \mathbf{X} $ are denoted by $ \mathbf{X}^\mathrm{T} $ and $ \mathbf{X}^\mathrm{H} $, respectively. Also, $ \mathbb{E} \left\lbrace \cdot \right\rbrace  $ denotes statistical expectation while $ \mathcal{CN} \left( \upsilon, \xi^2 \right) $ represents the complex Gaussian distribution with mean $ \upsilon $ and variance $ \xi^2 $. Symbols $ \left| \cdot \right| $ and $ \left\| \cdot \right\|_2 $ denote the absolute value and $ \ell_2 $-norm, respectively.


\section{System Model and Problem Formulation} \label{sec:system-model-problem-formulation}

\subsection{Preliminaries}
We consider an \gls{ISAC} system in which a full-duplex \gls{BS}, equipped with $ N_\mathrm{tx} $ transmit antennas and $ N_\mathrm{rx} $ receive antennas, serves a single-antenna user and senses a single target. The \gls{BS} operates in downlink mode and employs analog beamforming for transmission and reception. The transmit and receive beamformers can steer towards predefined angular directions, with multiple configurable beamwidths per direction. In addition, the \gls{BS} employs timeslot allocation to efficiently meet the communication and sensing demands over a finite horizon of $ S $ timeslots. Each timeslot can be exclusively assigned to either communication or sensing, or configured to support both functionalities concurrently. In particular, the objective of the \gls{RRM} design is to maximize communication throughput while ensuring that a specified number of sensing timeslots meet a predefined performance threshold. In the following, the $s$-th timeslot is denoted by $ \textsf{S}_s $, and the set of timeslots is defined as $ \mathcal{S} = \left\lbrace 1, \dots, S \right\rbrace $. The system model is depicted in \cref{fig:system-model}, with a toy example described below.

\textbf{Example:} \emph{During timeslots $ \textsf{S}_1 $ and $ \textsf{S}_2 $, the \gls{BS} employs a broad beam (orange) to simultaneously serve the user ($\mathsf{U}$) and illuminate the target ($\mathsf{T}$) for sensing. This is feasible due to their proximate angular positions. With a higher angular misalignment, the target would be served alone, thereby needing timeslots for only sensing. A dedicated receive beam (blue) is simultaneously steered toward the target to collect echo signals. In $\textsf{S}_3$ to $\textsf{S}_6$, the \gls{BS} switches to a narrower, high-gain beam (purple) directed exclusively at the user, since the target's sensing requirements have already been fulfilled in the first two timeslots (according to predefined requirements). This allows the \gls{BS} to concentrate the transmit power in the user's direction, thereby maximizing throughput.
}

\subsection{Timeslot allocation}
To determine which functionalities take place in each timeslot, we introduce constraints
\begin{align*}  
	& \red{\mathrm{C}_{1}}: \kappa_s = \left\lbrace 0, 1 \right\rbrace, \forall s \in \mathcal{S},   
	\\
	& \red{\mathrm{C}_{2}}: \zeta_s = \left\lbrace 0, 1 \right\rbrace, \forall s \in \mathcal{S},   
\end{align*}
where $ \kappa_s = 1 $ indicates that $ \textsf{S}_s $ is used for communication, and $ \kappa_s = 0 $ otherwise. Similarly, $ \zeta_s = 1 $ indicates that $ \textsf{S}_s $ is used for sensing, and $ \zeta_s = 0 $ otherwise.

To determine whether a given timeslot is active (i.e., used for communication, sensing, or both), we introduce constraint
\begin{align*}  
	\red{\mathrm{C}_{3}}: \gamma_s = \kappa_s \lor \zeta_s, \forall s \in \mathcal{S}. 
\end{align*}

Here, $ \gamma_s = 1 $ indicates that $ \textsf{S}_s $ is active, while $ \gamma_s = 0 $ indicates that $ \textsf{S}_s $ is idle. Furthermore, we include constraint
\begin{align*}  
	\red{\mathrm{C}_{4}}: \textstyle \sum_{ s \in \mathcal{S} } \gamma_s \leq S, 
\end{align*}
to ensure that the \gls{RRM} design is confined within a finite horizon of $ S $ timeslots.

\subsection{Beam selection} \label{sec:beam-selection}
The \gls{BS} can dynamically adjust the direction and beamwidth of its transmit and receive beampatterns, constrained to a finite set of possible configurations. We model this using a predefined codebook, where each codeword (i.e., codebook's element) corresponds to a unique combination of direction and beamwidth. Let $ {D}_\mathrm{tx} $ denote the different angular directions in which the transmit signal can be steered. For each direction, we assume $ {B}_\mathrm{tx} $ different beamwidths, leading to  $ L_\mathrm{tx} = D_\mathrm{tx} B_\mathrm{tx} $ distinct transmit beampatterns. We denote the transmit codewords as $ \mathbf{b}_b \in \mathbb{C}^{N_\mathrm{tx} \times 1} $, such that $ \big\| \mathbf{b}_b \big\|_2 = 1 $, where $ b \in \mathcal{L}_\mathrm{tx} $ and $ \mathcal{L}_\mathrm{tx} = \left\lbrace 1, 2, \dots, L_\mathrm{tx} \right\rbrace $. A similar approach is adopted with the receive codewords, considering $ D_\mathrm{rx} $ directions and $ B_\mathrm{rx} $ beamwidths. This leads to $ L_\mathrm{rx} = D_\mathrm{rx} B_\mathrm{rx} $ distinct receive beampattern possibilities, with the receive codewords denoted by $ \mathbf{c}_c \in \mathbb{C}^{N_\mathrm{rx} \times 1} $, where $ c \in \mathcal{L}_\mathrm{rx} $ and $ \mathcal{L}_\mathrm{rx} = \left\lbrace 1, 2, \dots, L_\mathrm{rx} \right\rbrace $.

To enable transmit beamforming, we introduce constraint
\begin{align*}  
	\red{\mathrm{C}_{5}}: \chi_{b,s} \in \left\lbrace 0, 1 \right\rbrace, \forall b \in \mathcal{L}_\mathrm{tx}, s \in \mathcal{S}, 
\end{align*}
where $ \chi_{b,s} = 1 $ indicates that codeword $ \mathbf{b}_b $ is used in $ \textsf{S}_s $, and $ \chi_{b,s} = 0 $ otherwise. Furthermore, we introduce constraint
\begin{align*}  
	\red{\mathrm{C}_{6}}: \textstyle \sum_{b \in \mathcal{L}_\mathrm{tx}} \chi_{b,s} = \gamma_s, \forall s \in \mathcal{S}, 
\end{align*}
to ensure that only one transmit codeword is selected for each active timeslot (see constraint $ \red{\mathrm{C}_{3}} $). Thus, the transmit beamformer employed in $ \textsf{S}_s $ is defined by constraint
\begin{align*}  
	\red{\mathrm{C}_{7}}: \mathbf{t}_s = \textstyle \sum_{b \in \mathcal{L}_\mathrm{tx}} \mathbf{b}_b \cdot \chi_{b,s}, \forall s \in \mathcal{S}.
\end{align*}

To enable receive beamforming, we introduce constraint
\begin{align*}  
	\red{\mathrm{C}_{8}}: \rho_{c,s} \in \left\lbrace 0, 1 \right\rbrace, \forall c \in \mathcal{L}_\mathrm{rx}, s \in \mathcal{S}, 
\end{align*}
where $ \rho_{c,s} = 1 $ indicates that codeword $ \mathbf{c}_c $ is used in $ \textsf{S}_s $, and  $ \rho_{c,s} = 0 $ otherwise. Additionally, we include constraint
\begin{align*}  
	\red{\mathrm{C}_{9}}: \textstyle \sum_{c \in \mathcal{L}_\mathrm{rx}} \rho_{c,s} = \zeta_s, \forall s \in \mathcal{S}, 
\end{align*}
to ensure that only one receive codeword is selected for each timeslot performing sensing (see constraint $ \red{\mathrm{C}_{2}} $). The receive beamformer used in $ \textsf{S}_s $ is given by constraint
\begin{align*}  
	\red{\mathrm{C}_{10}}: \mathbf{r}_s = \textstyle \sum_{c \in \mathcal{L}_\mathrm{rx}} \mathbf{c}_c \cdot \rho_{c,s}, \forall s \in \mathcal{S}. 
\end{align*}

\subsection{Communication metric}
The signal transmitted by the \gls{BS} in $ \textsf{S}_s $ is given by $ \mathbf{x}_s = \mathbf{t}_s d_s $, where $ d_s \in \mathbb{C} $ represents the symbol in $ \textsf{S}_s $, which is used for either or both functionalities simultaneously. Symbol $ d_s $ follows a complex Gaussian distribution with zero mean and unit variance, e.g., $ \mathbb{E} \left\lbrace d_s d_s^{*} \right\rbrace = 1 $. The signal received by the user in $ \textsf{S}_s $ is 
\begin{align}
	\begin{split}
	y_\mathrm{com}^{s} & = \mathbf{h}^\mathrm{H} \mathbf{x}_s \cdot \kappa_{s} + n_{\mathrm{com}} 
	 = \mathbf{h}^\mathrm{H} \mathbf{t}_s d_s \cdot \kappa_{s} + n_{\mathrm{com}}, 
	\end{split}
\end{align}
where $ \mathbf{h} \in \mathbb{C}^{N_\mathrm{tx} \times 1} $ is the communication channel between the \gls{BS} and the user, while $ n_{\mathrm{com}} \sim \mathcal{CN} \left( 0,\sigma_\mathrm{com}^2 \right) $ represents \gls{AWGN}. As a result, the communication \gls{SNR} in $ \textsf{S}_s $ is given by 
\begin{align*} 
	\mathsf{SNR}_\mathrm{com}^{s} = \left| \mathbf{h}^\mathrm{H} \mathbf{t}_s \cdot \kappa_{s} \right|^2 / \sigma_\mathrm{com}^2.
\end{align*}


\subsection{Sensing metric}
We model the target as a far-field point source and consider a monostatic radar configuration at the \gls{BS}, where the \gls{AOD} equals the \gls{AOA}, both denoted by $ \theta $. Additionally, the \gls{RC} captures both the target's radar cross-section and the round-trip path-loss relative to the \gls{BS}, and is denoted by $ \psi $. The sensing channel between the BS and the target is 
\begin{align} 
	\mathbf{G} = \psi \cdot \mathbf{a}_\mathrm{rx} \left( \theta \right) \mathbf{a}^\mathrm{H}_\mathrm{tx} \left( \theta \right)
	= \psi \cdot \mathbf{A} \left( \theta \right),
\end{align}
where the transmit and receive steering vectors are given by $ \mathbf{a}_l \left( \theta \right) = \tfrac{1}{\sqrt{N_l}} \mathrm{e}^{\mathrm{j} \boldsymbol{\phi}_l \cos \left( \theta \right)} $, with $ l = \left\lbrace \mathrm{tx}, \mathrm{rx} \right\rbrace $. Additionally, $ \boldsymbol{\phi}_l = \tfrac{2 \pi d_l}{\lambda} \left[  \tfrac{-N_l+1}{2}, \dots,  \tfrac{N_l-1}{2} \right]^\mathrm{T} $, $ \lambda$ is the wavelength, and $ d_l $ represents the spacing among antenna elements. 


The signal reflected by the target and received by the \gls{BS} during $ \mathsf{S}_s $ is modeled as
\begin{align} \label{eqn:signal-received-bs}
	\begin{aligned}
	y_\mathrm{sen}^{s} & = \mathbf{r}_s^\mathrm{H} \big( \mathbf{G} + \mathbf{Q} \big) \mathbf{x}_s \cdot \zeta_{s} + \mathbf{r}_s^\mathrm{H} \mathbf{n}_{\mathrm{sen}},
	\\
	& = \mathbf{r}_s^\mathrm{H} \mathbf{G} \mathbf{t}_s d_s \cdot \zeta_{s} + \mathbf{r}_s^\mathrm{H} \mathbf{Q}  \mathbf{t}_s d_s \cdot \zeta_{s} + \mathbf{r}_s^\mathrm{H} \mathbf{n}_{\mathrm{sen}},
	\end{aligned}
\end{align}
where $ \mathbf{Q} $ represents the direct \gls{SI} channel between the transmit and receive arrays, while $ \mathbf{n}_{\mathrm{sen}} \sim \mathcal{CN} \left( \mathbf{0}, \sigma_\mathrm{sen}^2 \mathbf{I} \right) $ denotes \gls{AWGN} at the \gls{BS}'s receiver. Specifically, $ \mathbf{Q} $ depends on the relative geometry between the arrays, and is modeled as 
\begin{align}
	\left[ \mathbf{Q} \right]_{\widebar{m},\widebar{n}} = \tfrac{\lambda}{4 \pi \widebar{d}_{\widebar{m},\widebar{n}}} \tfrac{1}{\sqrt{N_\mathrm{rx} N_\mathrm{tx}}} \mathrm{e}^{-\mathrm{j} \frac{2 \pi}{\lambda} \widebar{d}_{\widebar{m},\widebar{n}}},
\end{align}
where $ \widebar{d}_{\widebar{m},\widebar{n}} $ represents the distance between the $ \widebar{m} $-th receive antenna and the $ \widebar{n} $-th transmit antenna \cite{wang2023:near-field-integrated-sensing-communications}, with the array centers spaced by distance $ \widebar{d}_c $.

To mitigate \gls{SI} influence, the \gls{BS} employs analog and/or digital signal processing techniques \cite{kim2023:performance-analysis-self-interference-cancellation-full-duplex-massive-mimo-systems-subtraction-versus-spatial-suppression }. Leveraging an estimate of $ \mathbf{Q} $, denoted by $ \mathbf{Q}_\mathrm{est} = (1 - \upsilon_\mathrm{si} ) \mathbf{Q} $, the \gls{BS} subtracts $ \mathbf{r}_s^\mathrm{H} \mathbf{Q}_\mathrm{est} \mathbf{t}_s d_s \cdot \zeta_{s} $ from $ y_\mathrm{sen}^{s} $, which results in
\begin{align} \label{eqn:signal-received-bs-rsi}
\begin{aligned}
	\widebar{y}_\mathrm{sen}^{s} & = y_\mathrm{sen}^{s} - \mathbf{r}_s^\mathrm{H} \mathbf{Q}_\mathrm{est} \mathbf{t}_s d_s \cdot \zeta_{s},
	\\
	& = \mathbf{r}_s^\mathrm{H} \mathbf{G} \mathbf{t}_s d_s \cdot \zeta_{s} + \mathbf{r}_s^\mathrm{H} \mathbf{R} \mathbf{t}_s d_s \cdot \zeta_{s} + \mathbf{r}_s^\mathrm{H} \mathbf{n}_{\mathrm{sen}}.
\end{aligned}
\end{align}

Here, $ \mathbf{R} = \mathbf{Q} - \mathbf{Q}_\mathrm{est}  = \upsilon_\mathrm{si} \mathbf{Q} $ is the residual \gls{SI} channel, where $ \upsilon_\mathrm{si} $ captures the imperfection in the \gls{SI} cancellation process. A value of $ \upsilon_\mathrm{si} $ close to $ 1 $ indicates poor cancellation performance (i.e., high residual interference), whereas a value near $ 0 $ reflects highly effective cancellation. Since $ \upsilon_\mathrm{si} $ cannot be precisely known in practice, it is modeled as an uncertain parameter within a bounded interval
\begin{align*} 
	\red{\mathrm{C}_{11}}: \Upsilon \triangleq \left\lbrace \upsilon_\mathrm{si} \mid \upsilon_\mathrm{si} = \widebar{\upsilon}_\mathrm{si} + \Delta \upsilon_\mathrm{si}, \left| \Delta \upsilon_\mathrm{si} \right|^2 \leq \epsilon^2 \right\rbrace,
\end{align*}
imposing an infinite number of inequalities, where $ \upsilon_\mathrm{si} $ is the actual but unknown \gls{SI} factor, $ \widebar{\upsilon}_\mathrm{si} $ is the estimated \gls{SI} factor, and $ \Delta \upsilon_\mathrm{si} $ is random error whose power is bounded by $ \epsilon^2 $.


From (\ref{eqn:signal-received-bs-rsi}), the target's sensing \gls{SINR} during $ \mathsf{S}_s $ is defined as
\begin{align*} 
	\mathsf{SINR}_\mathrm{sen}^{s} =  
	 \left| \mathbf{r}_s^\mathrm{H}  \mathbf{G} \mathbf{t}_s \cdot \zeta_{s} \right|^2  / \left(  
	  \left|\mathbf{r}_s^\mathrm{H}  \mathbf{R} \mathbf{t}_s \cdot \zeta_{s} \right|^2 + \sigma_\mathrm{sen}^2 \left\| \mathbf{r}_s \right\|_2^2 \right).
\end{align*}

Subsequently, we introduce constraint
\begin{align*}
	\displaystyle \red{\mathrm{C}_{12}}: \mathsf{SINR}_\mathrm{sen}^{s} 
	\geq \Lambda_{\mathrm{sinr}} \cdot \zeta_{s}, \forall s \in \mathcal{S}, 
\end{align*}
which guarantees a predefined \gls{SINR} threshold $ \Lambda_{\mathrm{sinr}} $ for the target across its allocated timeslots. The minimum number of sensing timeslots is denoted by $ M_\mathrm{sen} $ and is ensured through
\begin{align*}
	 \red{\mathrm{C}_{13}}: \textstyle \sum_{s \in \mathcal{S}} \zeta_s \geq M_\mathrm{sen}.
\end{align*}

\subsection{Objective function}
We define the objective function 
\begin{align} 
	  f  (\boldsymbol{\Omega}) \triangleq \textstyle \sum_{ s \in \mathcal{S} } W T \log_2 \left( 1 + \mathsf{SNR}_\mathrm{com}^{s} \right),
\end{align}
which represents the throughput. Here, $ T $ is the timeslot duration, $ W $ is the system's bandwidth, and $ \boldsymbol{\Omega} $ represents the set of all decision variables.

\subsection{Problem formulation}
We formulate the \gls{RRM} design as
\begin{align*} 
	\mathcal{P} \left( \boldsymbol{\Omega} \right): & \underset{ \boldsymbol{\Omega} }{\mathrm{~maximize}}
	& f \left( \boldsymbol{\Omega} \right) ~~~ \mathrm{s.t.} ~~~ \boldsymbol{\Omega} \in \mathcal{X},
\end{align*}
where $ \mathcal{X} $ is the feasible domain defined by $ \red{\mathrm{C}_{1}} - \red{\mathrm{C}_{13}} $. Problem $ \mathcal{P} \left( \boldsymbol{\Omega} \right) $ is a semi-infinite nonconvex \gls{MINLP} that is challenging to solve due to variable couplings, its fractional structure, and an infinite number of constraints.


\section{Proposed Solution Approach} \label{sec:proposed-approach}

This section describes a sequence of transformation procedures applied to $ \mathcal{P} \left( \boldsymbol{\Omega} \right) $, resulting in an exact reformulation denoted by $ \mathcal{P}' \left( \boldsymbol{\Omega}' \right) $. Specifically, $ \boldsymbol{\Omega}' $ represents the updated set of variables associated with the transformed problem.

\begin{definition} \label{proc:procedure-1}

We substitute $ \mathbf{t}_s $ (as defined in $ \red{\mathrm{C}_{7}} $) into $ f (\boldsymbol{\Omega}) $, leading to 
\begin{align*}
f_{\mathrm{aux},1} (\boldsymbol{\Omega}) \triangleq \sum_{ s \in \mathcal{S} } W T \log_2 \big( 1 + \big| \sum_{b \in \mathcal{L}_\mathrm{tx}} \widebar{\mathbf{h}}^\mathrm{H} \mathbf{b}_b \cdot \chi_{b,s} \cdot \kappa_{s} \big|^2 \big), 
\end{align*}
where $ \widebar{\mathbf{h}} = \mathbf{h} / \sigma_\mathrm{com} $. Considering any $ \mathsf{S}_s $, and applying Jensen's inequality to the absolute value term in $ f_{\mathrm{aux},1} $, yields 
\begin{align*}
\red{\mathrm{D}_{\mathrm{aux},1}}: \big| \sum_{b \in \mathcal{L}_\mathrm{tx}} \widebar{\mathbf{h}}^\mathrm{H} \mathbf{b}_b \cdot \chi_{b,s} \cdot \kappa_s \big| \leq \sum_{b \in \mathcal{L}_\mathrm{tx}} \big| \widebar{\mathbf{h}}^\mathrm{H} \mathbf{b}_b \cdot \chi_{b,s} \cdot \kappa_s \big|.
\end{align*} 

If $ \mathbf{b}_{i} $ denotes the transmit beamformer in $ \mathsf{S}_s $, then $ \chi_{i,s} = 1 $ and $ \chi_{b,s} = 0 $,  $ \forall b \in \mathcal{L}_\mathrm{tx} \setminus \{i\} $. Hence, $ \red{\mathrm{D}_{\mathrm{aux},1}} $ simplifies to 
\begin{align*}
\red{\mathrm{D}_{\mathrm{aux},2}}: \big| \sum_{b \neq i} \widebar{\mathbf{h}}^\mathrm{H} \mathbf{b}_b \cdot \chi_{b,s} \cdot \kappa_s + \widebar{\mathbf{h}}^\mathrm{H} \mathbf{b}_{i} \cdot \chi_{i,s} \cdot \kappa_s \big| 
\\
\leq \sum_{b \neq i} \big| \widebar{\mathbf{h}}^\mathrm{H} \mathbf{b}_b \cdot \chi_{b,s} \cdot \kappa_s \big| + \big| \widebar{\mathbf{h}}^\mathrm{H} \mathbf{b}_{i} \cdot \chi_{i,s} \cdot \kappa_s \big|. 
\end{align*}

Exponentiating both sides yields 
\begin{align*}
\red{\mathrm{D}_{\mathrm{aux},3}}: \big| \sum_{b \neq i} \widebar{\mathbf{h}}^\mathrm{H} \mathbf{b}_b \cdot \chi_{b,s} \cdot \kappa_s + \widebar{\mathbf{h}}^\mathrm{H} \mathbf{b}_{i} \cdot \chi_{i,s} \cdot \kappa_s \big|^2 \leq 
\\
\big( \sum_{b \neq i} \big| \widebar{\mathbf{h}}^\mathrm{H} \mathbf{b}_b \cdot \chi_{b,s} \cdot \kappa_s \big| \big)^2 + \big| \widebar{\mathbf{h}}^\mathrm{H} \mathbf{b}_{i} \cdot \chi_{i,s} \cdot \kappa_s \big|^2 + 
\\
2 \big( \sum_{b \neq i} \big| \widebar{\mathbf{h}}^\mathrm{H} \mathbf{b}_b \cdot \chi_{b,s} \cdot \kappa_s \big| \big) \big| \widebar{\mathbf{h}}^\mathrm{H} \mathbf{b}_{i} \cdot \chi_{i,s} \cdot \kappa_s \big|.
\end{align*} 

The terms under the summations are zero, leading to $ \big| \sum_{b \in \mathcal{L}_\mathrm{tx}} \widebar{\mathbf{h}}^\mathrm{H} \mathbf{b}_b \cdot \chi_{b,s} \cdot \kappa_s \big|^2 = \sum_{b \in \mathcal{L}_\mathrm{tx}} \big| \widebar{\mathbf{h}}^\mathrm{H} \mathbf{b}_b \cdot \chi_{b,s} \cdot \kappa_s \big|^2 $ and allowing us to equivalently express $ f_{\mathrm{aux},1} (\boldsymbol{\Omega}) $ as 
\begin{align*}
f_{\mathrm{aux},2} (\boldsymbol{\Omega}) \triangleq \sum_{ s \in \mathcal{S} } W T \log_2 \big( 1 + \sum_{b \in \mathcal{L}_\mathrm{tx}} \big|  \widebar{\mathbf{h}}^\mathrm{H} \mathbf{b}_b \cdot \chi_{b,s} \cdot \kappa_{s} \big|^2 \big).
\end{align*} 

Leveraging the subadditivity property of the logarithm, we derive 
\begin{align*}
\red{\mathrm{D}_{\mathrm{aux},4}}: \log_2 \big( 1 + \sum_{b \in \mathcal{L}_\mathrm{tx}} \big|  \widebar{\mathbf{h}}^\mathrm{H} \mathbf{b}_b \cdot \chi_{b,s} \cdot \kappa_{s} \big|^2 \big) \leq 
\\
\sum_{b \in \mathcal{L}_\mathrm{tx}} \log_2 \big( 1 + \big|  \widebar{\mathbf{h}}^\mathrm{H} \mathbf{b}_b \cdot \chi_{b,s} \cdot \kappa_{s} \big|^2 \big).
\end{align*} 

Recalling that $ \chi_{i,s} = 1 $ and $ \chi_{b,s} = 0 $,  $ \forall b \in \mathcal{L}_\mathrm{tx} \setminus \{i\} $, we expand both sides of $ \red{\mathrm{D}_{\mathrm{aux},4}} $ and eliminate the zero-valued terms, as we did in $ \red{\mathrm{D}_{\mathrm{aux},2}} $. This reveals that the inequality in $ \red{\mathrm{D}_{\mathrm{aux},4}} $ is tight, allowing us to recast  $ f_{\mathrm{aux},2} (\boldsymbol{\Omega}) $ as 
\begin{align*}
f' (\boldsymbol{\Omega}') \triangleq \sum_{s \in \mathcal{S}} \sum_{b \in \mathcal{L}_\mathrm{tx}} W T \log_2 \big( 1 + \big| \widebar{\mathbf{h}}^\mathrm{H} \mathbf{b}_b \big|^2 \delta_{b,s} \big),
\end{align*}
where $ \delta_{b,s} $ are newly introduced variables, defined in 
\begin{align*}
\red{\mathrm{D}_{\mathrm{aux},5}}: \delta_{b,s} = \chi_{b,s} \kappa_s, \forall b \in \mathcal{L}_\mathrm{tx}, s \in \mathcal{S}. 
\end{align*}

To cope with the coupling between variables $ \chi_{b,s} $ and $ \kappa_s $ within $ \red{\mathrm{D}_{\mathrm{aux},5}} $, we use propositional calculus as in \cite{abanto2024:radio-resource-management-design-RSMA-optimization-beamforming-user-admission-discrete-continuous-rates-imperfect-sic, abanto2024:optimal-user-target-scheduling-user-target-pairing-low-resolution-phase-only-beamforming-isac-systems}, which decomposes $ \red{\mathrm{D}_{\mathrm{aux},5}} $ into
\begin{align*}
& \red{\mathrm{D}_{1}}: \delta_{b,s} \leq \chi_{b,s}, \forall b \in \mathcal{L}_\mathrm{tx}, s \in \mathcal{S}, 
\\
& \red{\mathrm{D}_{2}}: \delta_{b,s} \leq \kappa_s, \forall b \in \mathcal{L}_\mathrm{tx}, s \in \mathcal{S}, 
\\
&
\red{\mathrm{D}_{3}}: \delta_{b,s} \geq \chi_{b,s} + \kappa_s - 1, \forall b \in \mathcal{L}_\mathrm{tx}, s \in \mathcal{S},
\\
&
\red{\mathrm{D}_{4}}: \delta_{b,s} \geq 0, \forall b \in \mathcal{L}_\mathrm{tx}, s \in \mathcal{S}.
\end{align*}

As a result, $ f (\boldsymbol{\Omega}) $ is equivalently rewritten as $ f' (\boldsymbol{\Omega}') $, subject to including extra constraints $ \red{\mathrm{D}_{1}} $, $ \red{\mathrm{D}_{2}} $, $ \red{\mathrm{D}_{3}} $, and $ \red{\mathrm{D}_{4}} $.
\end{definition}

\begin{definition} \label{proc:procedure-2}

By rearranging the terms of $ \red{\mathrm{C}_{12}} $, this constraint can be transformed into 
\begin{align*}
\red{\mathrm{E}_{\mathrm{aux},1}}: \big| \mathbf{r}_s^\mathrm{H}  \mathbf{G} \mathbf{t}_s \cdot \zeta_{s} \big|^2 \geq \zeta_s \Lambda_{\mathrm{sinr}} \big|\mathbf{r}_s^\mathrm{H}  \mathbf{R} \mathbf{t}_s \big|^2 + 
\\
\zeta_s \Lambda_{\mathrm{sinr}} \sigma_\mathrm{sen}^2 \left\| \mathbf{r}_s \right\|_2^2, \forall s \in \mathcal{S}.
\end{align*} 

Since $ \zeta_{s} $ is binary and common to all terms in $ \red{\mathrm{E}_{\mathrm{aux},1}} $, it can be canceled from both sides\footnote{Note that canceling $ \zeta_s $ from both sides of $ \red{\mathrm{E}_{\mathrm{aux},1}} $ is generally invalid if $ \zeta_s = 0 $. However, $ \mathbf{r}_s $ is a function of $ \zeta_s $, (through $ \red{\mathrm{C}_{9}} $ and $ \red{\mathrm{C}_{10}} $). Thus, if sensing is enabled, $ \mathbf{r}_s \neq \mathbf{0} $ (as one codeword must be selected), which implies $ \zeta_s = 1 $ and justifies the cancellation. Conversely, if sensing is not enabled, then $ \mathbf{r}_s = \mathbf{0} $ (through $ \red{\mathrm{C}_{9}} $ and $ \red{\mathrm{C}_{10}} $), reducing the inequality to $ 0 = 0 $, which holds true as an equality. Hence, in this particular case where $ \mathbf{r}_s $ retains information about $ \zeta_s $, the variables $ \zeta_s $ can be canceled on both sides. }, transforming $ \red{\mathrm{E}_{\mathrm{aux},1}} $ into 
\begin{align*}
\red{\mathrm{E}_{\mathrm{aux},2}}: \big| \mathbf{r}_s^\mathrm{H}  \mathbf{G} \mathbf{t}_s \big|^2 \geq \Lambda_{\mathrm{sinr}} \big|\mathbf{r}_s^\mathrm{H} \mathbf{R} \mathbf{t}_s \big|^2 + 
\\
\Lambda_{\mathrm{sinr}} \sigma_\mathrm{sen}^2 \left\| \mathbf{r}_s \right\|_2^2, \forall s \in \mathcal{S}. 
\end{align*} 

Leveraging $ \red{\mathrm{C}_{11}} $ and the definitions $ \mathbf{G} = \psi \mathbf{A} \left( \theta \right) $ and $ \mathbf{R} = \upsilon_\mathrm{si} \mathbf{Q} $, we can reformulate $ \red{\mathrm{E}_{\mathrm{aux},2}} $ as 
\begin{align*}
\red{\mathrm{E}_{\mathrm{aux},3}}: \big| \mathbf{r}_s^\mathrm{H} \mathbf{A} \left( \theta \right) \mathbf{t}_s \big|^2 \geq \max_{ \upsilon_\mathrm{si} \in \Upsilon } \frac{\Lambda_{\mathrm{sinr}}}{\left| \psi \right|^2} \upsilon_\mathrm{si}^2 \big|\mathbf{r}_s^\mathrm{H}  \mathbf{Q} \mathbf{t}_s \big|^2 +
\\
\frac{\Lambda_{\mathrm{sinr}}}{\left| \psi \right|^2} \sigma_\mathrm{sen}^2 \left\| \mathbf{r}_s \right\|_2^2, \forall s \in \mathcal{S}.
\end{align*}  

To improve tractability, we introduce a new set of variables, defined by 
\begin{align*}
	\red{\mathrm{E}_{1}}: z_s \geq 0, \forall s \in \mathcal{S},
\end{align*}
allowing us to split $ \red{\mathrm{E}_{\mathrm{aux},3}} $ into $ \red{\mathrm{E}_{2}} $ and $ \red{\mathrm{E}_{\mathrm{aux},4}} $, shown below,
\begin{align*}
	& \red{\mathrm{E}_{2}}: \big| \mathbf{r}_s^\mathrm{H} \mathbf{A} \left( \theta \right) \mathbf{t}_s \big|^2  \geq z_s, \forall s \in \mathcal{S},
	\\
	& \red{\mathrm{E}_{\mathrm{aux},4}}: z_s \geq \max_{ \upsilon_\mathrm{si} \in \Upsilon } \frac{\Lambda_{\mathrm{sinr}}}{\left| \psi \right|^2} \upsilon_\mathrm{si}^2 \big|\mathbf{r}_s^\mathrm{H}  \mathbf{Q} \mathbf{t}_s \big|^2 + 
	\\
	& 
	~~~~~~~~~~~~~~ \frac{\Lambda_{\mathrm{sinr}}}{\left| \psi \right|^2} \sigma_\mathrm{sen}^2  \left\| \mathbf{r}_s \right\|_2^2, \forall s \in \mathcal{S}, \forall s \in \mathcal{S}. 
\end{align*}

Note that $ \red{\mathrm{E}_{\mathrm{aux},4}} $ is equivalent to 
\begin{align*}
\red{\mathrm{E}_{\mathrm{aux},5}}: z_s \geq \max_{ \left| \Delta \upsilon_\mathrm{si} \right|^2 \leq \epsilon^2 }  \frac{\Lambda_{\mathrm{sinr}}}{\left| \psi \right|^2} \left( \widebar{\upsilon}_\mathrm{si} + \Delta \upsilon_\mathrm{si} \right)^2  \big|\mathbf{r}_s^\mathrm{H}  \mathbf{Q} \mathbf{t}_s \big|^2 +
\\
\frac{\Lambda_{\mathrm{sinr}}}{\left| \psi \right|^2} \sigma_\mathrm{sen}^2 \left\| \mathbf{r}_s \right\|_2^2, \forall s \in \mathcal{S}, \forall s \in \mathcal{S}, 
\end{align*}
if we leverage $ \red{\mathrm{C}_{11}} $. The maximum value of the \gls{RHS} term in $ \red{\mathrm{E}_{\mathrm{aux},5}} $ is achieved when $ \Delta \upsilon_\mathrm{si} $ has a positive sign and maximum magnitude, i.e., $ \Delta \upsilon_\mathrm{si} = \epsilon $. Hence, we can recast $ \red{\mathrm{E}_{\mathrm{aux},5}} $ as 
\begin{align*}
	\red{\mathrm{E}_{3}}: z_s \geq \frac{\Lambda_{\mathrm{sinr}}}{\left| \psi \right|^2} \left(  \left( \widebar{\upsilon}_\mathrm{si} + \epsilon \right)^2 \big|\mathbf{r}_s^\mathrm{H}  \mathbf{Q} \mathbf{t}_s \big|^2 + \sigma_\mathrm{sen}^2 \left\| \mathbf{r}_s \right\|_2^2\right) , \forall s \in \mathcal{S}, 
\end{align*}

As a result, constraints $ \red{\mathrm{C}_{11}} $ and $ \red{\mathrm{C}_{12}} $ can be recast equivalently as $ \red{\mathrm{E}_{1}} $, $ \red{\mathrm{E}_{2}} $, and $ \red{\mathrm{E}_{3}} $.

\end{definition}

\begin{definition} \label{proc:procedure-3}
 
By applying propositional calculus as in \cite{abanto2024:radio-resource-management-design-RSMA-optimization-beamforming-user-admission-discrete-continuous-rates-imperfect-sic}, we eliminate the logical OR ($ \lor $) in constraint $ \red{\mathrm{C}_{3}} $, thereby transforming it into four equivalent constraints,
\begin{align*}
	& \red{\mathrm{F}_{1}}: \gamma_s \leq \kappa_s + \zeta_s, \forall s \in \mathcal{S},
	\\
	& \red{\mathrm{F}_{2}}: \gamma_s \geq \kappa_s, \forall s \in \mathcal{S},
	\\
	& \red{\mathrm{F}_{3}}: \gamma_s \geq \zeta_s, \forall s \in \mathcal{S},
	\\
	& \red{\mathrm{F}_{4}}: \gamma_s \leq 1, \forall s \in \mathcal{S}. 
\end{align*}

Consequently,  $ \red{\mathrm{C}_{3}} $ is equivalent to $ \red{\mathrm{F}_{1}} $, $ \red{\mathrm{F}_{2}} $, $ \red{\mathrm{F}_{3}} $, and $ \red{\mathrm{F}_{4}} $.

\end{definition}

\begin{definition} \label{proc:procedure-4}
	Substituting $ \mathbf{r}_s $ and $ \mathbf{t}_s $ (as defined in $ \red{\mathrm{C}_{7}} $ and $ \red{\mathrm{C}_{10}} $) into $ \red{\mathrm{E}_{2}} $ leads to 
	\begin{align*}
	\red{\mathrm{G}_{\mathrm{aux},1}}: \big| \sum_{b \in \mathcal{L}_\mathrm{tx}} \sum_{c \in \mathcal{L}_\mathrm{rx}} \mathbf{c}_c^\mathrm{H} \mathbf{A} \left( \theta \right) \mathbf{b}_b \rho_{c,s} \chi_{b,s} \big|^2 \geq z_s, \forall s \in \mathcal{S}.
	\end{align*} 
	
	Considering any $ \mathsf{S}_s $ and applying Jensen's inequality to the absolute value term in $ \red{\mathrm{G}_{\mathrm{aux},1}} $, results in 
	\begin{align*}
		\red{\mathrm{G}_{\mathrm{aux},2}}: \big| \sum_{b \in \mathcal{L}_\mathrm{tx}} \sum_{c \in \mathcal{L}_\mathrm{rx}} \mathbf{c}_c^\mathrm{H} \mathbf{A} \left( \theta \right) \mathbf{b}_b \rho_{c,s} \chi_{b,s} \big| \leq 
		\\
		\sum_{b \in \mathcal{L}_\mathrm{tx}} \sum_{c \in \mathcal{L}_\mathrm{rx}} \big| \mathbf{c}_c^\mathrm{H} \mathbf{A} \left( \theta \right) \mathbf{b}_b \rho_{c,s} \chi_{b,s} \big|.
	\end{align*} 

	Assuming that $ \mathbf{b}_i $ and $ \mathbf{c}_j $ denote the transmit and receive beams used in $ \mathsf{S}_s $ leads to $ \chi_{i,s} = 1 $, $ \rho_{j,s} = 1 $, $ \chi_{b,s} = 0, \forall b \in \mathcal{L}_\mathrm{tx} \setminus \{i\} $, and $ \rho_{c,s} = 0, \forall c \in \mathcal{L}_\mathrm{rx} \setminus \{j\} $. As a result, $ \red{\mathrm{G}_{\mathrm{aux},2}} $ collapses to 
	\begin{align*}
	\red{\mathrm{G}_{\mathrm{aux},3}}: \big| \sum_{(b,c) \neq (i,j)} \mathbf{c}_c^\mathrm{H} \mathbf{A} \left( \theta \right) \mathbf{b}_b \rho_{c,s} \chi_{b,s} + \mathbf{c}_j^\mathrm{H} \mathbf{A} \left( \theta \right) \mathbf{b}_i \rho_{j,s} \chi_{i,s} \big| 
	\\
	\leq  \sum_{(b,c) \neq (i,j)} \big| \mathbf{c}_c^\mathrm{H} \mathbf{A} \left( \theta \right) \mathbf{b}_b \rho_{c,s} \chi_{b,s} \big| + \big| \mathbf{c}_j^\mathrm{H} \mathbf{A} \left( \theta \right) \mathbf{b}_i \rho_{j,s} \chi_{i,s} \big|. 
	\end{align*} 
	
	Exponentiating both sides to the power of two and eliminating the zero-valued terms, as in \textbf{Procedure~1}, reveals that both sides of the inequality are identical. As a result, the absolute value and summation can be interchanged without affecting the result, allowing us to equivalently express $ \red{\mathrm{G}_{\mathrm{aux},1}} $ as 
	\begin{align*}
	\red{\mathrm{G}_{\mathrm{aux},4}}: \sum_{b \in \mathcal{L}_\mathrm{tx}} \sum_{c \in \mathcal{L}_\mathrm{rx}} \big| \mathbf{c}_c^\mathrm{H} \mathbf{A} \left( \theta \right) \mathbf{b}_b \rho_{c,s} \chi_{b,s} \big|^2 \geq z_s, \forall s \in \mathcal{S}. 
	\end{align*} 
	
	To eliminate the coupling among $ \rho_{c,s} $ and $ \chi_{b,s} $ in $ \red{\mathrm{G}_{\mathrm{aux},4}} $, we introduce new variables, defined in 
	\begin{align*}
	\red{\mathrm{G}_{\mathrm{aux},5}}: \pi_{b,c,s} = \rho_{c,s} \chi_{b,s}, \forall b \in \mathcal{L}_\mathrm{tx}, c \in \mathcal{L}_\mathrm{rx}, s \in \mathcal{S}, 
	\end{align*}
	which enables us to transform $ \red{\mathrm{G}_{\mathrm{aux},4}} $ into 
	\begin{align*}
		\red{\mathrm{G}_{1}}: \sum_{b \in \mathcal{L}_\mathrm{tx}} \sum_{c \in \mathcal{L}_\mathrm{rx}} \big| \mathbf{c}_c^\mathrm{H} \mathbf{A} \left( \theta \right) \mathbf{b}_b \pi_{b,c,s} \big|^2 \geq z_s, \forall s \in \mathcal{S}. 
	\end{align*}
	Leveraging propositional calculus, we decompose $ \red{\mathrm{G}_{\mathrm{aux},5}} $ into 
	\begin{align*}
		& \red{\mathrm{G}_{2}}: \pi_{b,c,s} \leq \chi_{b,s}, \forall b \in \mathcal{L}_\mathrm{tx}, c \in \mathcal{L}_\mathrm{rx}, s \in \mathcal{S},
		\\
		& \red{\mathrm{G}_{3}}: \pi_{b,c,s} \leq \rho_{c,s}, \forall b \in \mathcal{L}_\mathrm{tx}, c \in \mathcal{L}_\mathrm{rx}, s \in \mathcal{S},
		\\
		& \red{\mathrm{G}_{4}}: \pi_{b,c,s} \geq \chi_{b,s} + \rho_{c,s} - 1, \forall b \in \mathcal{L}_\mathrm{tx}, c \in \mathcal{L}_\mathrm{rx}, s \in \mathcal{S},
		\\
		& \red{\mathrm{G}_{5}}: \pi_{b,c,s} \geq 0, \forall b \in \mathcal{L}_\mathrm{tx}, c \in \mathcal{L}_\mathrm{rx}, s \in \mathcal{S}.
	\end{align*}
	as in \cite{abanto2024:radio-resource-management-design-RSMA-optimization-beamforming-user-admission-discrete-continuous-rates-imperfect-sic}.
	Following a similar procedure as that used to convert $ \red{\mathrm{G}_{\mathrm{aux},1}} $ into $ \red{\mathrm{G}_{\mathrm{aux},4}} $, we can transform $ \red{\mathrm{E}_{3}} $ into 
	\begin{align*}
	\red{\mathrm{G}_{\mathrm{aux},6}}: z_s \geq \frac{\Lambda_{\mathrm{sinr}}}{\left| \psi \right|^2} \left( \widebar{\upsilon}_\mathrm{si} + \epsilon \right)^2 \sum_{b \in \mathcal{L}_\mathrm{tx}} \sum_{c \in \mathcal{L}_\mathrm{rx}} \big| \mathbf{c}_c^\mathrm{H} \mathbf{Q} \mathbf{b}_b \pi_{b,c,s} \big|^2 +
	\\
	\frac{\Lambda_{\mathrm{sinr}}}{\left| \psi \right|^2} \sigma_\mathrm{sen}^2 \left\| \mathbf{r}_s \right\|_2^2, \forall s \in \mathcal{S}, \forall s \in \mathcal{S}, 
	\end{align*}
	where the new variables $ \pi_{b,c,s} $, introduced in $ \red{\mathrm{G}_{\mathrm{aux},5}} $, have been adopted. Note that $ \left\| \mathbf{r}_s \right\|_2^2 $, which appears in $ \red{\mathrm{G}_{\mathrm{aux},6}} $, can be transformed by leveraging $ \red{\mathrm{C}_{10}} $, leading to 
	\begin{align*}
		\red{\mathrm{G}_{\mathrm{aux},7}}: \sum_{c' \in \mathcal{L}_\mathrm{rx}} \sum_{c \in \mathcal{L}_\mathrm{rx}} \mathbf{c}^\mathrm{H}_{c'} \mathbf{c}_c \cdot \rho_{c',s} \cdot \rho_{c,s}.
	\end{align*}
	
	Since only one receive beam is chosen per active timeslot, which we denote by $ \mathbf{c}_j $, then the product $ \rho_{c',s} \cdot \rho_{c,s} $ is zero unless $ j = c = c' $. Thus, $ \left\| \mathbf{r}_s \right\|_2^2 = \sum_{c \in \mathcal{L}_\mathrm{rx}} \mathbf{c}^\mathrm{H}_{c} \mathbf{c}_c \cdot \rho_{c,s} = \sum_{c \in \mathcal{L}_\mathrm{rx}} \left\| \mathbf{c}_c \right\|_2^2 \cdot \rho_{c,s} $. Using this result, allows us to recast $ \red{\mathrm{G}_{\mathrm{aux},7}} $ as 
	\begin{align*}
		 \red{\mathrm{G}_{6}}:  z_s \geq & \frac{\Lambda_{\mathrm{sinr}}}{\left| \psi \right|^2} \left( \widebar{\upsilon}_\mathrm{si} + \epsilon \right)^2 \sum_{b \in \mathcal{L}_\mathrm{tx}} \sum_{c \in \mathcal{L}_\mathrm{rx}} \big| \mathbf{c}_c^\mathrm{H} \mathbf{Q} \mathbf{b}_b \pi_{b,c,s} \big|^2 +
		\\ 
		& \frac{\Lambda_{\mathrm{sinr}}}{\left| \psi \right|^2} \sigma_\mathrm{sen}^2 \sum_{c \in \mathcal{L}_\mathrm{rx}} \left\| \mathbf{c}_c \right\|_2^2 \cdot \rho_{c,s}, \forall s \in \mathcal{S}.
	\end{align*}

	Thus, $ \red{\mathrm{C}_{7}} $, $ \red{\mathrm{C}_{10}} $, $ \red{\mathrm{C}_{12}} $, $ \red{\mathrm{E}_{2}} $, and $ \red{\mathrm{E}_{3}} $ are equivalently transformed into $ \red{\mathrm{G}_{1}} $, $ \red{\mathrm{G}_{2}} $, $ \red{\mathrm{G}_{3}} $, $ \red{\mathrm{G}_{4}} $, $ \red{\mathrm{G}_{5}} $, and $ \red{\mathrm{G}_{6}} $.
	
\end{definition}

The reformulated problem is expressed as
\begin{align*} 
	\mathcal{P}' \left( \boldsymbol{\Omega}' \right): & \underset{ \boldsymbol{\Omega'} }{\mathrm{~maximize}}
	& f' \left( \boldsymbol{\Omega}' \right) ~~~ \mathrm{s.t.} ~~~ \boldsymbol{\Omega}' \in \mathcal{X}',
\end{align*}
where $ \mathcal{X}' $ is the feasible domain defined by $ \red{\mathrm{C}_{1}} $, $ \red{\mathrm{C}_{2}} $, $ \red{\mathrm{C}_{4}} - \red{\mathrm{C}_{6}} $, $ \red{\mathrm{C}_{8}} $, $ \red{\mathrm{C}_{9}} $, $ \red{\mathrm{C}_{13}} $, $ \red{\mathrm{D}_{1}} - \red{\mathrm{D}_{4}} $, $ \red{\mathrm{E}_{1}} $, $ \red{\mathrm{F}_{1}} - \red{\mathrm{F}_{4}} $, and $ \red{\mathrm{G}_{1}} - \red{\mathrm{G}_{6}} $. Here, $ \mathcal{P}' \left( \boldsymbol{\Omega}' \right) $ is a \gls{MILP} which can be solved to global optimality. 

\begin{remark}
The worst-case computational complexity of solving $\mathcal{P}' \left( \boldsymbol{\Omega}' \right)$ is given by $\binom{S}{M_\mathrm{sen}} L_\mathrm{rx}^{M_\mathrm{sen}} L_\mathrm{tx}^{S - M_\mathrm{sen}}$. However, due to the assumed channel invariance over the time horizon of $S$ timeslots, we can impose that the first $M_\mathrm{sen}$ timeslots are allocated for sensing and the remaining timeslots for communication. This allows the complexity to be reduced to $L_\mathrm{rx}^{M_\mathrm{sen}} L_\mathrm{tx}^{S - M_\mathrm{sen}}$ without affecting optimality. However, in practice, the computational complexity is significantly lower because the problem is a \gls{MILP}. These types of problems can be solved more efficiently using specialized off-the-shelf solvers that employ advanced techniques like branch-and-bound and cutting planes.
\end{remark}


\section{Simulation Results} \label{sec:simulation-results}

We evaluate problem $\mathcal{P}' \left( \boldsymbol{\Omega}' \right)$ under various configurations using the Rician channel model. The communication channel is $ \mathbf{h} = \gamma \mathbf{v} $, where $ \gamma $ accounts for large-scale fading and
\begin{equation}
	\mathbf{v} = \sqrt{K / (K+1)} \mathbf{v}^\mathrm{LoS} + \sqrt{1/(K+1)} \mathbf{v}^\mathrm{NLoS},
\end{equation}
is the normalized small-scale fading, with $ K = 100 $ being the Rician fading factor. The \gls{LoS} component is given by $ \mathbf{v}^\mathrm{LoS} = \tfrac{1}{\sqrt{N_\mathrm{tx}}} \mathrm{e}^{\mathrm{j} \boldsymbol{\phi}_\mathrm{tx} \cos \left( \beta \right)} $, where $ \beta $ is the \gls{LoS} angle, and the \gls{NLoS} components are defined as $ \mathbf{v}^\mathrm{NLoS} \sim \mathcal{CN} \left( \mathbf{0}, \mathbf{I} \right) $. For large-scale fading, we adopt the \texttt{UMa} channel model \cite{3gpp:38.901}, modeled as $ \gamma = 28 + 22 \log_{10} (l) + 20 \log_{10} (f_\mathrm{c}) $ dB, where $ f_\mathrm{c} = 41 $ GHz is the carrier frequency, and $ l = 60 $ m is the distance between the \gls{BS} and the user. 

The communication and sensing noise powers are $ \sigma_\mathrm{com}^2 = -114 $ dB and $ \sigma_\mathrm{sen}^2 = -74 $ dB, respectively. The bandwidth is $ W = 200 $ MHz and the timeslot duration is $ T = 1 $ ms. The \gls{BS} is equipped with \glspl{ULA} consisting of $ N_\mathrm{tx} = 8 $ transmit antennas and $ N_\mathrm{rx} = 16 $ receive antennas. The transmit and receive directions span the interval from $ 50^\circ $ to $ 130^\circ $ with spacing of $ 5^\circ $ degrees, yielding $ D_\mathrm{tx} = 17 $ and $ D_\mathrm{rx} = 17 $ distinct directions for transmission and reception, respectively\footnote{In this work, the codewords are generated using the discrete Fourier transform. To broaden the beamwidth, the extreme elements of the arrays are switched off while the power is increased proportionally to maintain the same total transmit or receive power across codewords, as per the design condition specified in \cref{sec:beam-selection}.}. The beamwidths for transmission are $ \left\lbrace 13^\circ, 17^\circ, 26^\circ, 60^\circ \right\rbrace $, while the beamwidth options for reception are $ \left\lbrace  6^\circ, 13^\circ, 17^\circ, 26^\circ \right\rbrace $, leading to $ B_\mathrm{tx} = 4 $ and $ B_\mathrm{rx} = 4 $. The transmit and receive powers are set to $ 1 $ W and $ 0.25 $ W, respectively. The target's \gls{RC} is $ \psi = 6 \cdot 10^{-4} $ and the distance between array centers is $ \widebar{d}_\mathrm{c} = 0.15 $ m. The results are averaged over $ 50 $ realizations unless specified otherwise.

\begin{figure*}[!t]
 	\begin{subfigure}[b]{0.19\textwidth}
		\begin{center}
			\includegraphics[]{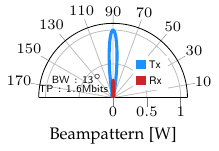}
			\caption{$ \theta = 90 \mid \Lambda_{\mathrm{sinr}} = 3 $}
			\label{fig:results-scenario-1a}
		\end{center}
 	\end{subfigure}
    \hfill 
 	\begin{subfigure}[b]{0.19\textwidth}
		\begin{center}
			\includegraphics[]{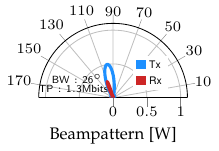}
			\caption{$ \theta = 110 \mid \Lambda_{\mathrm{sinr}} = 3 $}
			\label{fig:results-scenario-1b}
		\end{center}
 	\end{subfigure}
    \hfill 
 	\begin{subfigure}[b]{0.19\textwidth}
		\begin{center}
			\includegraphics[]{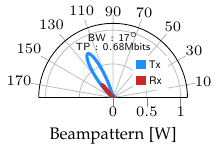}
			\caption{$ \theta = 130 \mid \Lambda_{\mathrm{sinr}} = 3 $}
			\label{fig:results-scenario-1c}
		\end{center}
 	\end{subfigure}
 	\hfill
 	\begin{subfigure}[b]{0.19\textwidth}
		\begin{center}
			\includegraphics[]{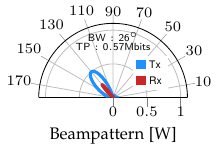}
			\caption{$ \theta = 130 \mid \Lambda_{\mathrm{sinr}} = 4 $}
			\label{fig:results-scenario-1d}
		\end{center}
 	\end{subfigure}
 	\hfill
 	\begin{subfigure}[b]{0.19\textwidth}
		\begin{center}
			\includegraphics[]{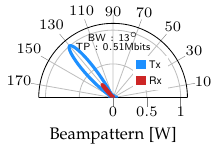}
			\caption{$ \theta = 130 \mid \Lambda_{\mathrm{sinr}} = 5 $}
			\label{fig:results-scenario-1e}
		\end{center}
 	\end{subfigure}
    \caption{Impact of user and target alignment on beam adaptation. }
 	\label{fig:results-scenario-1}
 	\vspace{-4mm}
\end{figure*}

\subsection{Scenario I}

We investigate how the transmit beam adapts to the relative angular positions of the user and target. To facilitate a clear visualization of this behavior, \cref{fig:results-scenario-1} considers a scenario with a single timeslot, thereby enforcing shared usage between sensing and communication (i.e., $ S = S_\mathrm{sen} = 1 $). We assume the user remains fixed at $ \beta = 90^\circ $, while the target moves across $ \theta = \left\lbrace 90^\circ, 110^\circ, 130^\circ \right\rbrace $. Throughout \cref{fig:results-scenario-1a} to \cref{fig:results-scenario-1c}, we set $ \Lambda_{\mathrm{sinr}} = 3 $, while in \cref{fig:results-scenario-1d} and \cref{fig:results-scenario-1e}, we consider $ \Lambda_{\mathrm{sinr}} = 4 $ and $ \Lambda_{\mathrm{sinr}} = 5 $, respectively. Additionally, perfect \gls{SI} cancellation is assumed ($ \widebar{\upsilon}_\mathrm{si} = 0 $ and $ \epsilon = 0 $).

In \cref{fig:results-scenario-1a}, the user and target are perfectly aligned ($ \theta = \beta = 90^\circ $), enabling the \gls{BS} to employ the narrowest transmit beamwidth of $ 13^\circ $ to service both. This results in a high throughput of $ 1.6 $ Mbits. In \cref{fig:results-scenario-1b}, under moderate misalignment, the transmit beam steers towards $ 100^\circ $ to approach the target direction (at $ \theta = 110^\circ $) and widens its beamwidth to $ 26^\circ $ to cover both user and target. As a result, the throughput decreases to $ 1.3 $ Mbits. In \cref{fig:results-scenario-1c}, with greater misalignment ($ \theta = 130^\circ $), the beam is steered further towards the target at $ 125^\circ $, to guarantee $ \Lambda_{\mathrm{sinr}} = 3 $, but the throughput degrades significantly to $ 0.68 $ Mbits due to reduced energy radiated towards the user. In \cref{fig:results-scenario-1d}, the sensing threshold increases to $ \Lambda_{\mathrm{sinr}} = 4 $, prompting the beam to be centered at the target's \gls{AOD} ($ \theta = 130^\circ $) and to widen its beamwidth to $ 26^\circ $ to ensure some energy also reaches the user, resulting in a reduced throughput of $ 0.57 $ Mbits. In \cref{fig:results-scenario-1e}, with an even more stringent threshold of $ \Lambda_{\mathrm{sinr}} = 5 $, the beam narrows back to $ 13^\circ $, concentrating primarily on the target. This tighter focus leads to a further drop in throughput to $ 0.51 $ Mbits, as only a minimal portion of the beam energy reaches the user. In all cases, the receive beam consistently points towards the target's \gls{AOD} utilizing the narrowest available beamwidth of $ 6^\circ $ to maximize sensing accuracy, as no \gls{SI} is present.

\subsection{Scenario II}

\begin{figure}[!h]
		\vspace{-2mm}
		\begin{center}
			\includegraphics[]{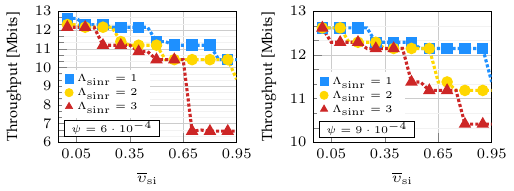}
			\vspace{-2mm}
    		\caption{Impact of residual SI and sensing threshold. }
 			\label{fig:results-scenario-3a}
 		\end{center}
 		\vspace{-3mm}
\end{figure}

We investigate how residual \gls{SI}, the sensing \gls{SINR} threshold, and the target's \gls{RC} influence the throughput performance. The results are illustrated in \cref{fig:results-scenario-3a}, where we consider $ \theta = 100^\circ $, $ \Lambda_{\mathrm{sinr}} \in \left\lbrace 1, 2, 3 \right\rbrace $, $ S = 8 $, $ S_\mathrm{sen} = 4 $, $ \widebar{\upsilon}_\mathrm{si} \in \left[ 0, 0.95 \right] $, $ \epsilon = 0.05 $, and $ \psi = \left\lbrace 6 \cdot 10^{-4}, 9 \cdot 10^{-4} \right\rbrace $.

As $ \widebar{\upsilon}_\mathrm{si} $ increases, the throughput consistently decreases across all \gls{SINR} thresholds. This degradation is attributed to elevated residual \gls{SI}, which diminishes the effective sensing \gls{SINR}. To compensate, the \gls{BS} allocates more directive power towards the target, consequently reducing the energy radiated to the user. Additionally, increasing the sensing \gls{SINR} threshold $ \Lambda_{\mathrm{sinr}} $ also reduces throughput, particularly when $ \widebar{\upsilon}_\mathrm{si} $ is moderate to high. This is due to the need to steer more transmit power towards the target's \gls{AOD}. Comparing the two \gls{RC} scenarios, we observe that a larger \gls{RC} (i.e., $ \psi = 9 \cdot 10^{-4} $) has a favorable impact on throughput. Specifically, a higher \gls{RC} enhances the sensing \gls{SINR} without requiring the beam to be tightly aligned with the target's \gls{AOD}. This allows more transmit power to be directed towards the user, thereby mitigating the throughput degradation caused by high \gls{SINR} thresholds or elevated residual \gls{SI}.

\subsection{Scenario III}

We investigate how the throughput is affected by the residual \gls{SI} and the number of timeslots allocated for sensing. We adopt the same parameter settings as in \textit{Scenario~II}, with $ \Lambda_{\mathrm{sinr}} = 3 $, $ S_\mathrm{sen} \in \left\lbrace 1, \dots, 8 \right\rbrace $, and $ \psi = 6 \cdot 10^{-4} $. The throughput is visualized as a heatmap in \cref{fig:results-scenario-3a}.

\begin{figure}[!h]
		\vspace{-2mm}
		\begin{center}
			\includegraphics[]{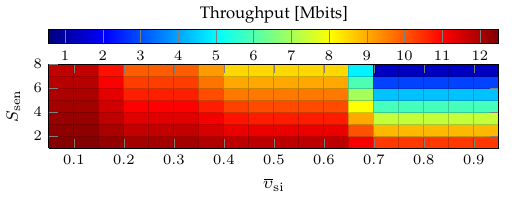}
			\vspace{-2mm}
    		\caption{Impact of residual SI and sensing timeslots. }
 			\label{fig:results-scenario-3b}
 		\end{center}
 		\vspace{-5mm}
\end{figure}

For any fixed value of $ S_\mathrm{sen} $, the throughput consistently decreases as $ \widebar{\upsilon}_\mathrm{si} $ increases. This is because higher $ \widebar{\upsilon}_\mathrm{si} $ exacerbates the effect of residual \gls{SI} on the sensing \gls{SINR}, forcing the \gls{BS} to employ increasingly directional beams steered towards the target's \gls{AOD} to maintain sensing performance, a configuration that inherently reduces communication throughput. Additionally, for any fixed $ \widebar{\upsilon}_\mathrm{si} $, throughput declines as $ S_\mathrm{sen} $ increases, since more stringent sensing requirements constrain the resources available for throughput maximization.  In contrast, neglecting residual \gls{SI} significantly reduces reliability, yielding only $ 46.2\% $ average feasibility over all cases, as shown in \cref{fig:results-scenario-3b}. This substantial performance gap underscores the necessity of explicitly incorporating residual \gls{SI} into the \gls{RRM} design to ensure robust operation.

\subsection{Scenario IV}

\begin{figure}[!h]
		\vspace{-2mm}
		\begin{center}
			\includegraphics[]{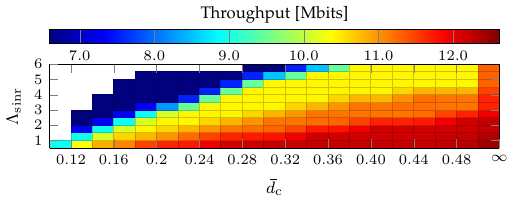}
			\vspace{-2mm}
    		\caption{Impact of array separation.}
 			\label{fig:results-scenario-4}
 		\end{center}
 		\vspace{-5mm}
\end{figure}

We investigate how the throughput is influenced by the separation between the transmit and receive arrays, which directly affects the severity of \gls{SI}. We adopt the same parameter settings as in \textit{Scenario~III}, with $ S_\mathrm{sen} = 4 $, and assume that no active \gls{SI} cancellation is employed. Instead, \gls{SI} is mitigated solely through physical array separation. The resulting throughput is depicted as a heatmap in \cref{fig:results-scenario-4}.

The results show that increasing the separation distance $ \widebar{d}_\mathrm{c} $ has a beneficial impact on throughput. As $ \widebar{d}_\mathrm{c} $ grows, the influence of \gls{SI} diminishes, allowing the \gls{BS} to employ narrower, more directive beams towards the target's \gls{AOD} without the risk of overwhelming \gls{SI} power. Conversely, increasing the sensing \gls{SINR} threshold $ \Lambda_{\mathrm{sinr}} $ leads to a reduction in throughput, as more transmit power must be dedicated to meet the sensing requirement. The white regions indicate infeasible allocations where the \gls{SI} is too severe to satisfy the sensing \gls{SINR} threshold. We observe that a separation distance of $ 50 $ cm is sufficient to mitigate the impact of \gls{SI} to acceptable levels. In particular, at this distance, the resulting throughput remains within $ 12\% $ of the ideal case with hypothetically infinite separation, indicating that a distance of $ 50 $ cm offers a practical balance between \gls{SI} cancellation and transceiver footprint. 

\section{Conclusions} \label{sec:conclusions}

This paper investigated a novel \gls{RRM} problem, jointly addressing timeslot allocation and beam adaptation under practical considerations, including discrete beam directions and beamwidths, and imperfect \gls{SI} cancellation. To tackle the inherent complexity of the \gls{RRM} problem, we proposed a tractable solution by reformulating the problem as a \gls{MILP}, thereby ensuring reliable performance even under uncertainty in the residual \gls{SI} level. Our results revealed how user-target angular alignment significantly influences system performance, with beam adaptation serving as a key mechanism to balance sensing and communication demands. Furthermore, we examined the effects of residual \gls{SI} on throughput, and showed that increasing the physical separation between transmit and receive arrays offers a means of mitigating \gls{SI}.


\section*{Acknowledgment} \label{sec:acknowledgment}

The authors acknowledge the financial support by the Federal Ministry for Research, Technology and Space (BMFTR) in Germany in the programme of “Souverän. Digital. Vernetzt.” Joint project 6G-RIC, project identification number: 16KISK035.

\bibliographystyle{IEEEtran}
\bibliography{IEEEabrv,ref}


\begin{appendices}

\renewcommand{\thesectiondis}[2]{\Alph{section}:}

\end{appendices}

\end{document}